\RequirePackage{mathtools}
\RequirePackage{commath, amsmath, amssymb, amsfonts}
\documentclass[10pt, a4paper]{article}
\usepackage[margin=1in]{geometry}
\usepackage{authblk}
\usepackage{float}

\usepackage[backend=bibtex, sorting=none, autopunct=false, biblabel=brackets, url=false, chaptertitle=false, style=phys]{biblatex}
\usepackage{graphicx}
\graphicspath{{./figs/}}
\usepackage{hyperref}
\hypersetup{
	colorlinks   = true, 
	urlcolor     = blue, 
	linkcolor    = blue, 
	citecolor   = blue 
}

\newcommand{\ie}{i.e.~}
\newcommand{\cf}{c.f.~}
\newcommand{\eg}{e.g.~}

\newcommand{\va}[1]{\mathbf{#1}}

\newcommand{\ket}[1]{|{#1}\rangle}

\newcommand{\braket}[2]{\langle{#1}|{#2}\rangle}
\newcommand{\mel}[3]{\langle{#1}|{#2}|{#3}\rangle}
\newcommand{\field}[1]{\boldsymbol{\mathcal{{#1}}}}
\newcommand{\vdot}{\boldsymbol\cdot}

\newcommand{\pop}{\va{\hat{p}}}
\newcommand{\rop}{\va{\hat{r}}}
\newcommand{\jop}{\va{\hat{j}}}
\newcommand{\efield}{\mathcal{F}}

\begin{document}

\title{Two-dimensional Materials with Giant Optical Nonlinearities Near the Theoretical Upper Limit}

\author[1,2,3]{Alireza Taghizadeh\thanks{ata@nano.aau.dk}}
\author[3,4]{Kristian S. Thygesen}
\author[1,2]{Thomas G. Pedersen}
\affil[1]{Department of Materials and Production, Aalborg University, 9220 Aalborg {\O}st, Denmark}
\affil[2]{Center for Nanostructured Graphene (CNG), 9220 Aalborg {\O}st, Denmark}
\affil[3]{Computational Atomic-scale Materials Design (CAMD), Department of Physics, Technical University of Denmark (DTU), 2800 Kgs. Lyngby, Denmark}
\affil[4]{Center for Nanostructured Graphene (CNG), Technical University of Denmark (DTU), 2800 Kgs. Lyngby, Denmark}
\makeatletter
\renewcommand*{\@fnsymbol}[1]{\ensuremath{\ifcase#1\or \dagger\or *\or \ddagger\or
    \mathsection\or \mathparagraph\or \|\or **\or \dagger\dagger
    \or \ddagger\ddagger \else\@ctrerr\fi}}
\makeatother
\maketitle

\begin{abstract}
Nonlinear optical (NLO) phenomena such as harmonic generation, Kerr, and Pockels effects are of great technological importance for lasers, frequency converters, modulators, switches, etc. Recently, two-dimensional (2D) materials have drawn significant attention due to their strong and unique NLO properties. Here, we describe an efficient first-principles workflow for calculating the quadratic optical response and apply it to 375 non-centrosymmetric semiconductor monolayers from the Computational 2D Materials Database (C2DB). Sorting the non-resonant nonlinearities with respect to bandgap $E_g$ reveals an upper limit proportional to $E_g^{-4}$, which is neatly explained by two distinct generic models. We identify multiple promising candidates with giant nonlinearities and bandgaps ranging from 0.4 to 5 eV, some of which approach the theoretical upper limit and greatly outperform known materials. Our comprehensive library of ab initio NLO spectra for all 375 monolayers is freely available via the C2DB website. We expect this work to pave the way for highly efficient and compact opto-electronic devices based on 2D materials. 	
\end{abstract}


\section*{Introduction \label{sec:Introduction}}

\noindent Nonlinear optical phenomena relying on strong light-matter interactions in intense laser fields are the basis for numerous photonics and opto-electronics technologies with applications in telecommunication \cite{Schneider2004}, light sources \cite{Byer2000, Popmintchev2010}, bio-science \cite{Zipfel2003, Campagnola2003}, data storage \cite{Garmire2013}, and sensors \cite{Mesch2016}. In addition, nonlinear optical (NLO) spectroscopy is extensively employed to obtain insight into materials properties \cite{Prylepa2018} that are not accessible by \eg linear optical spectroscopy. For instance, polarization-resolved second-harmonic generation (SHG) is used to identify crystal orientations and symmetries \cite{Li2013, Wu2014}. SHG has also been used for direct visualization of the electric field inside integrated circuits \cite{Chen2003} or examining fabricated nanostructures with high spatial and temporal resolutions \cite{Jarrett2015}. 
Recently, two-dimensional (2D) materials and their van der Waals heterostructures \cite{Geim2013, Liu2016} have become attractive for realizing compact NLO devices due to their extreme thinness and strong NLO response \cite{Liu2016b, Hafez2018, Autere2018, Jiang2018}. In particular, highly nonlinear 2D materials have already been integrated into photonic devices \cite{Sun2016, Autere2018}. Moreover, their atomic thickness and relaxed phase-matching restrictions in the out-of-plane direction \cite{Wang2017} make 2D materials uniquely suited for novel applications such as nonlinear quantum cavity optics \cite{Wild2018}. The strong NLO effects add to a long list of unique and readily-tunable optical, electronic, and mechanical properties exhibited by 2D materials such as graphene \cite{Novoselov2004}, hexagonal boron nitride \cite{Tran2015}, phosphorene \cite{Liu2014, Xia2014}, transition metal dichalcogenides (TMDs) (\eg MoS$_2$, WS$_2$) \cite{Mak2010, Manzeli2017} as well as group-IV monochalcogenides (\eg GeSe, SnS), see Ref.~\cite{Ferrari2015} for a comprehensive review. 




While several studies have suggested that atomically thin 2D materials can exhibit stronger NLO response (per unit volume) than bulk materials \cite{Kumar2013, Dasgupta2019}, questions like ``what is the fundamental upper limit to the NLO response of 2D materials" and  ``how close are known 2D materials to the limit" remain unanswered to the best of our knowledge. Answering these questions are critical to gauge the true NLO potential of 2D materials and harness it via rational design of new materials that could form the basis for future nonlinear devices. Since many opto-electronic devices rely on low-loss propagation of light, the correlation of nonlinearity with bandgap $E_g$ (which determines the transparent frequency regime) is of particular importance \cite{Tan2019, Cook2017}. 

Due to the complexity of higher-order optical processes and nonlinear measurement techniques, theory and computations play an essential role in interpretion of NLO experiments. In particular, ab initio calculations have been successfully employed to explain magnitude and frequency dependence of nonlinear susceptibilities in bulk semiconductors yielding valuable microscopic insight \cite{Sharma2004, Nastos2005, Hbener2010}. Such ab initio calculations, in particular when conducted in a high-throughput fashion, can aid and potentially even lead the discovery of new materials with strong nonlinear response \cite{Kuzyk2000, Tan2019}. In the realm of 2D materials, however, NLO ab initio calculations have so far only been performed for a handful of the most popular materials such as MoS$_2$ \cite{Wang2015}, WS$_2$ \cite{Janisch2014}, WSe$_2$ \cite{Wang2015b}, hBN \cite{Kim2013}, Ge$_2$Se$_2$ \cite{Wang2017}, and $\alpha$-Bi \cite{Guo2020}. 


\begin{figure}[t]
	\centering
	\includegraphics[width=0.9\textwidth]{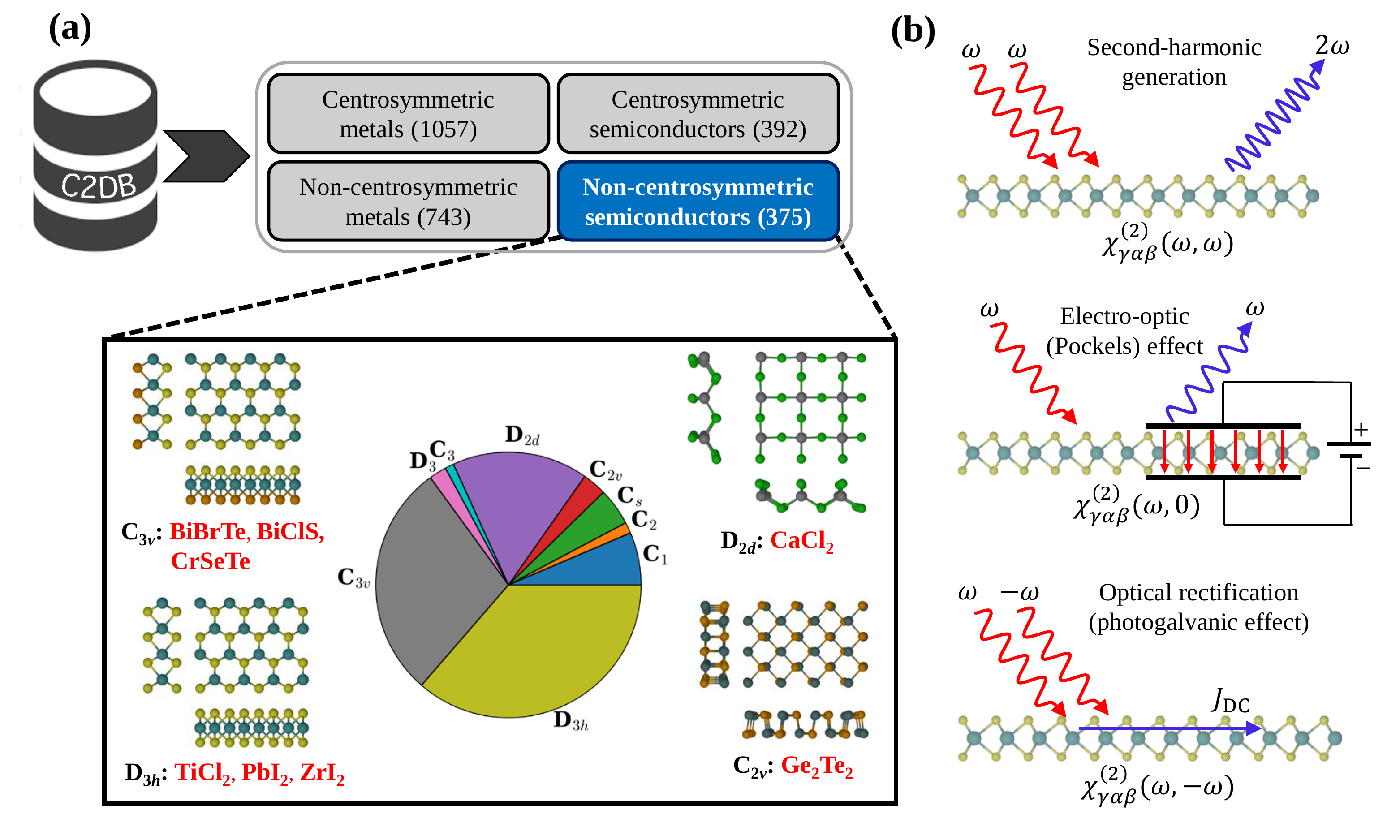}
	\caption[Schematic]{\textbf{Schematic of material selection and important quadratic optical processes.} (a) C2DB includes thousands of 2D materials with various crystal structures and chemical compositions. We screen non-magnetic non-centrosymmetric semiconductors from the C2DB for large optical nonlinearities. The numbers in parenthesis count the available monolayers in each category. Their distribution according to point group is shown in the pie chart along with examples of promising monolayers with strong quadratic optical response. (b) Schematic views of three important quadratic processes: (top) second-harmonic generation, (middle) electro-optic (Pockels) effect, and (bottom) optical rectification. For each process, the relevant susceptibility tensor is specified. } 
	\label{fig:schematic}
\end{figure}


In this work, we perform ab initio high-throughput computation of the quadratic optical response of 375 (non-centrosymmetric and non-magnetic) 2D semiconductors selected from the open Computational 2D Materials Database (C2DB) \cite{Haastrup2018}. By analyzing the DC limit of the quadratic response, \ie the non-resonant quadratic susceptibility, we identify several promising materials with extremely high nonlinearities exceeding those of the known champion materials WS$_2$ and Ge$_2$Se$_2$. We derive an analytical upper bound for the non-resonant quadratic susceptibility proportional to $E_g^{-4}$, which is closely followed by the most promising candidates identified in our ab initio data search. Intriguingly, the scaling of this upper limit with energy gap is similar to but distinct from that of molecular systems \cite{Kuzyk2000}. As another example of a useful correlation, we show that the NLO response is often strong in compounds containing Cr, Ti, or Te. The full frequency-dependent SHG spectra of all 375 materials are available from the C2DB \cite{C2DB} where they can be browsed or downloaded for further analysis. 






\section*{Results}


\subsection*{Quadratic Optical Response Theory \label{sec:Theory}}
Here, we briefly review the microscopic theory of the second-order optical response of a periodic system under the influence of an external electromagnetic field in the dipole approximation (long-wavelength regime). A general electric field $\field{F}$ can be decomposed into its harmonic components and written as
\begin{align}
&\field{F}(t) = \sum_{\alpha,\omega_1} \efield_\alpha(\omega_1) \va{e}_\alpha e^{-i\omega_1 t}  \, ,
\end{align}
where $\omega_1$ runs over positive and negative frequencies, $\va{e}_\alpha$ with $\alpha\in\{x,y,z\}$ denotes the unit vector along the $\alpha$-direction, and $\efield_\alpha(\omega_1)$ is the electric field at frequency of $\omega_1$. Note that, in the dipole approximation, we neglect local field effects, \ie the electric field does not vary spatially. The electric field induces a polarization, $\field{P}(t)$, that is given by $\field{P}(t)=\sum_{N=1}^{\infty} \field{P}^{(N)}(t)$, where $\field{P}^{(N)}$ is proportional to the $N$th power of the electric field, and the proportionality factor is the $N$th-order susceptibility tensor. Starting from the first-order term, the usual susceptibility tensor, linked to the dielectric constant and linear absorption, is determined from $\mathcal{P}_\gamma^{(1)}(t)=\epsilon_0 \sum_{\alpha,\omega_1} \chi_{\gamma\alpha}^{(1)}(\omega_1) \efield_\alpha(\omega_1) e^{-i\omega_1 t}$ ($\epsilon_0$ denotes the vacuum permittivity). Going beyond the linear regime, the quadratic susceptibility tensor, $\chi_{\gamma\alpha\beta}^{(2)}$, is defined via
\begin{align}
&\mathcal{P}_\gamma^{(2)}(t) = \epsilon_0 \sum_{\omega_1\omega_2} \sum_{\alpha\beta} \chi_{\gamma\alpha\beta}^{(2)}(\omega_1,\omega_2) \efield_\alpha(\omega_1) \efield_\beta(\omega_2) e^{-i(\omega_1+\omega_2) t} \, .
\end{align}
Here, $\chi_{\gamma\alpha\beta}^{(2)}$ is a rank-3 tensor with at most 18 independent elements due to intrinsic permutation symmetry, \ie $\chi_{\gamma\alpha\beta}^{(2)}(\omega_1,\omega_2)=\chi_{\gamma\beta\alpha}^{(2)}(\omega_2,\omega_1)$. Furthermore, the point group symmetry restricts the number of independent tensor elements and their relationships due to the Neumann principle \cite{Boyd2008}.

In the dipole approximation, $\chi_{\gamma\alpha\beta}^{(2)}$ vanishes identically irrespective of frequencies for centrosymmetric materials, while in non-centrosymmetric systems, several distinct quadratic processes are described by $\chi_{\gamma\alpha\beta}^{(2)}$. Three important quadratic processes with their corresponding susceptibilities are illustrated schematically in Fig.~\ref{fig:schematic}(b). In the SHG process, described by $\chi_{\gamma\alpha\beta}^{(2)}(\omega,\omega)$, two incident photons at frequency $\omega$ generate an emitted photon at frequency of $2\omega$. The Pockels effect is a quadratic process, in which the dielectric constant of the material is tuned via a static electric field and, hence, is essential for realizing electro-optic modulators \cite{Sun2016, Yu2017}. Computationally, the effect corresponds to the limit $\chi_{\gamma\alpha\beta}^{(2)}(\omega,0)$, where one photon frequency approaches zero, while the other is kept finite. In the optical rectification process, a DC current is generated from two incident photons, which leads to the well-known shift and injection currents in semiconductors \cite{Sipe2000, Cook2017, Wang2019}.
In the majority of opto-electronic devices, one would like to avoid significant light absorption and, hence, devices operate at frequencies well below the semiconductor bandgap (the so-called non-resonant regime).
Therefore, the static limit of the quadratic susceptibility, in which both frequencies approach zero, is of great importance for device applications. In this limit, the susceptibilities of all three processes in Fig.~\ref{fig:schematic}(b) for a given material converge to the same value. In addition, Kleinman symmetry is obeyed meaning that all permutations among $\gamma\alpha\beta$ are equal in this limit \cite{Boyd2008}.  

A general expression for $\chi_{\gamma\alpha\beta}^{(2)}(\omega_1,\omega_2)$ containing only microscopic quantities such as transition dipole moments and energy levels can be obtained using perturbation theory and density matrices. Previously \cite{Aversa1995, Sharma2004, Nastos2005, Wang2017, Taghizadeh2017}, the microscopic SHG susceptibility $\chi_{\gamma\alpha\beta}^{(2)}(\omega,\omega)$ has been studied  as a special case of the general quadratic response.  
In the present work, several distinct frequency combinations are considered and we, therefore, briefly review the derivation of the general quadratic susceptibility for two arbitrary frequencies in the method section. For a semiconductor, the derivation demonstrates that the general expression for $\chi_{\gamma\alpha\beta}^{(2)}(\omega_1,\omega_2)$ in the so-called length gauge has contributions from both purely interband ($e$) and mixed inter-intraband ($i$) processes, \ie $\chi_{\gamma\alpha\beta}^{(2)}(\omega_1,\omega_2) = \chi_{\gamma\alpha\beta}^{(2e)}(\omega_1,\omega_2)  + \chi_{\gamma\alpha\beta}^{(2i)}(\omega_1,\omega_2)$, and their expressions are given in Eqs.~(\ref{eq:chiinter}) and (\ref{eq:chiintra}), respectively. The purely interband term originates from the change in band populations (diagonal terms of the density matrix), whereas the mixed inter-intraband contribution is due to coherences (off-diagonal terms of the density matrix), see Ref.~\cite{Taghizadeh2019b} for details. 


Our calculations are performed in the independent particle approximation (IPA), based solely on Kohn-Sham quasi-particle states obtained from density functional theory (DFT). Thus, the Coulomb interaction between electrons and holes, \ie excitonic effects, and the quasi-particle self-energy corrections are neglected. Excitonic effects are pronounced in 2D materials due to quantum confinement and weak dielectric screening \cite{Wang2018} and, hence, important for accurate calculations of both linear \cite{Qiu2013, Ugeda2014, Thygesen2017} and NLO responses \cite{Gruning2014, Trolle2015, Taghizadeh2018, Taghizadeh2019} of 2D materials, particularly for large-bandgap materials. Excitons usually manifest themselves as a redshift of the optical response onset and significant transfer of spectral weight to bound excitons \cite{Qiu2013, Taghizadeh2018, Hipolito2018}. In the context of DFT, self-energy corrections and electron-hole interactions should ideally be included within the GW and Bethe-Salpeter equation (BSE) frameworks \cite{Qiu2013, Ugeda2014, Gruning2014, Wang2015b, Attaccalite2019}. These types of calculations are enormously computationally demanding, but provide better agreement with experimental results \cite{Qiu2013, Wang2015b, Thygesen2017}. Nonetheless, the non-resonant quadratic response is relatively insensitive to self-energy corrections and excitonic effects \cite{Pedersen2007}. For instance, the DC limits of the quadratic response in monolayer MoS$_2$ with or without excitons agree within a factor of 2 \cite{Taghizadeh2019b}. This should be contrasted with the material-specific variations among the systems considered in the current study, which spread over three orders of magnitude. Furthermore, the dielectric screening of the surroundings, \eg due to the substrate, reduces the excitonic effects in 2D materials \cite{Latini2015}. Therefore, frequently, the order of magnitude for the NLO response is accurately captured in the IPA, \eg see Refs.~\cite{Gruning2014, Attaccalite2019}.
Importantly, the calculated DFT bandgap typically agrees surprisingly well with the experimental optical bandgap, which is partly due to the cancellation between the self-energy correction and exciton binding energy \cite{Komsa2012, Janisch2014}. Hence, an IPA-based search may be utilized to identify promising nonlinear materials, which then can subsequently be investigated more accurately within the GW and BSE frameworks, if greater precision is required.  


\subsection*{Computational Overview}
We extract the relaxed crystal structures of non-magnetic and non-centrosymmetric semiconductors from the C2DB. The electronic band energies and wavefunctions are obtained from a DFT-PBE \cite{Perdew1996, Enkovaara2010} ground state calculation. Next, momentum and position matrix elements are calculated and stored. Finally, based on the crystal point-group symmetry, all non-zero elements of the quadratic susceptibility are identified and computed. The key feature of our approach is that the calculations can be automatized, allowing one to perform thousands of calculations in parallel without human intervention. The details of the methodology are provided in the method section. For simplicity, we neglect the spin-orbit coupling in our calculations, since it barely affects the low-frequency quadratic responses.
Note that the bulk susceptibility (with SI units of m/V) is ill-defined for 2D materials, since the volume cannot be defined without ambiguity in 2D systems. Instead, the sheet susceptibility, obtained by replacing volume with area and expressed in unit of m$^2$/V, is an unambiguous quantity for 2D materials. In practice, we place the 2D material between two large vacuum regions, and a full periodic structure (in three dimensions) with an artificial periodicity along the out-of-plane direction ($z$-direction) is simulated. The vacuum regions should be sufficiently wide that unphysical coupling between neighboring cells is avoided. Then, the bulk susceptibility is calculated for this structure using Eqs.~(\ref{eq:chiinter}) and (\ref{eq:chiintra}). Finally, the result is transformed to the unambiguous sheet susceptibility by multiplying with the width of the unit cell in the $z$-direction. 

  
\begin{figure}[t]
	\centering
	\includegraphics[width=0.9\textwidth]{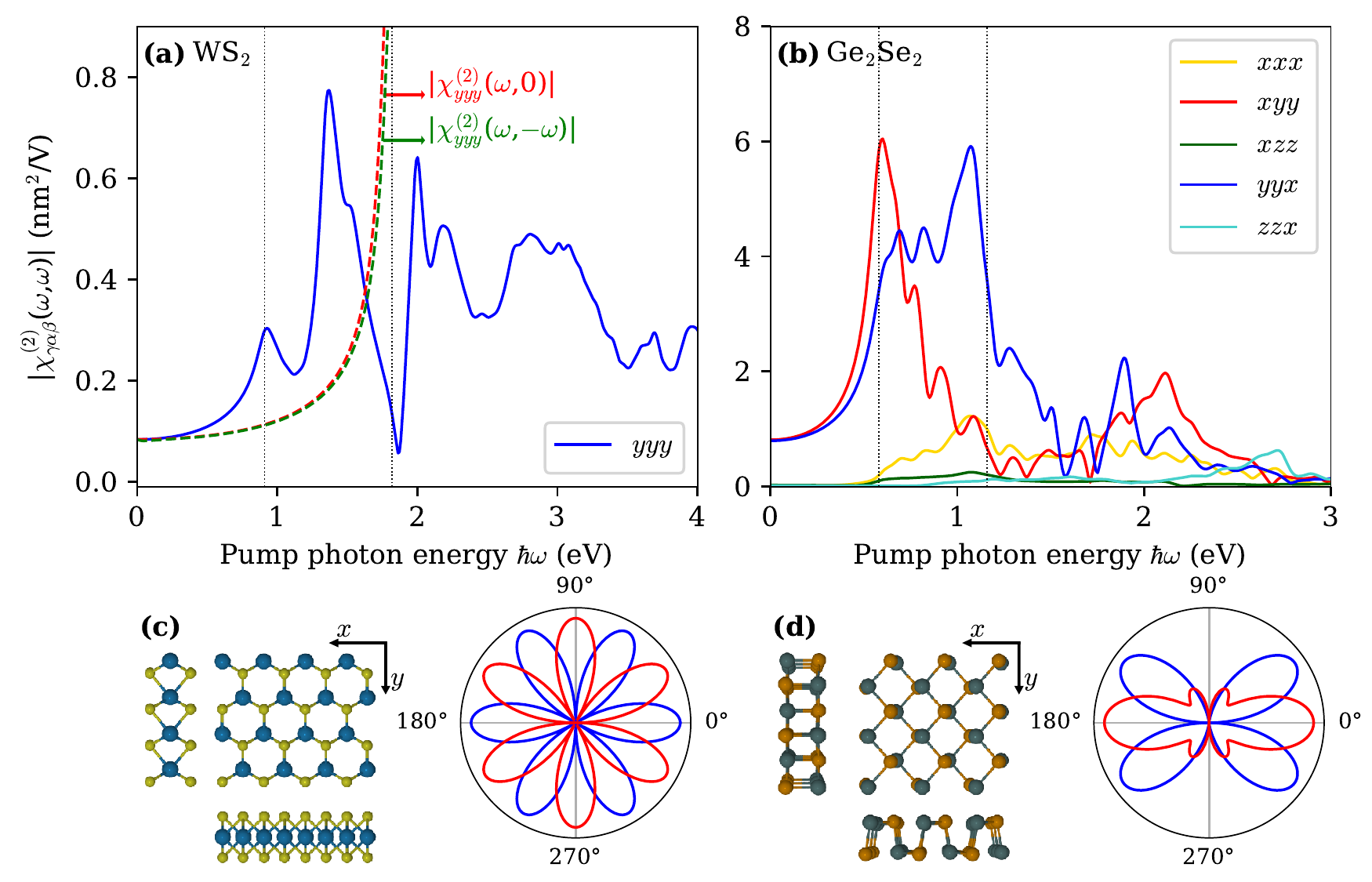}
	\caption[Spectra]{\textbf{Quadratic response of monolayer WS$_2$ and Ge$_2$Se$_2$.} SHG spectra of monolayer (a) WS$_2$ and (b) Ge$_2$Se$_2$, where only non-vanishing independent tensor elements are shown. The vertical dotted lines mark $\hbar\omega=E_g/2$ and $\hbar\omega=E_g$. In (a), quadratic susceptibilities describing the electro-optic effect $\chi_{yyy}^{(2)}(\omega,0)$ and optical rectification $\chi_{yyy}^{(2)}(\omega,-\omega)$ are included as red and green dashed lines, respectively (only below the bandgap). Crystal structures including top view and cross sectional views are shown in the left panel for (c) WS$_2$ and (d) Ge$_2$Se$_2$. The rotational anisotropy of SHG for parallel (blue) and perpendicular (red) configurations are illustrated next to the crystal structures with angles defined with respect to the crystal $x$-axis.   } 
	\label{fig:spectra}
\end{figure}

\subsection*{Benchmarking the Methodology}
We first discuss our results for two well-known and important nonlinear 2D materials: monolayer WS$_2$ and Ge$_2$Se$_2$. The former is a TMD with a hexagonal crystal structure (space group 187 and point group D$_{3h}$), while the latter is a group-IV dichalcogenide with an orthorhombic crystal structure (space group 31 and point group C$_{2v}$), see Figs.~\ref{fig:spectra}(c) and \ref{fig:spectra}(d). Consequently, monolayer WS$_2$ has only one independent tensor element, $\chi_{yyy}^{(2)}=-\chi_{yxx}^{(2)}=-\chi_{xyx}^{(2)}=-\chi_{xxy}^{(2)}$, whereas monolayer Ge$_2$Se$_2$ has 5 independent tensor elements, $\chi_{xxx}^{(2)}$, $\chi_{xyy}^{(2)}$, $\chi_{xzz}^{(2)}$, $\chi_{yyx}^{(2)}=\chi_{yxy}^{(2)}$, and $\chi_{zzx}^{(2)}=\chi_{zxz}^{(2)}$. Note that, throughout our study, we have inspected susceptibility tensor symmetries for various crystal classes, and confirmed the expected symmetry relations by direct calculation. We illustrate the magnitude of the SHG susceptibility versus frequency for monolayer WS$_2$ and Ge$_2$Se$_2$ in Figs.~\ref{fig:spectra}(a) and \ref{fig:spectra}(b), respectively. Both spectra are in good agreement with previously reported results for the two monolayers in Refs.~\cite{He2019} and \cite{Wang2015}, respectively. 
As mentioned above, for a given material, the DC limits of all quadratic responses are identical. This fact is validated here for the case of WS$_2$, for which we have computed the spectra (below the bandgap) of susceptibility tensors corresponding to the in-plane electro-optic and optical-rectification effects, \ie $\chi_{yyy}^{(2)}(\omega,0)$ and $\chi_{yyy}^{(2)}(\omega,-\omega)$, respectively, as shown by the dashed curves in Fig.~\ref{fig:spectra}(a). 
At photon energies well below the bandgap, we have $\chi_{\gamma\alpha\beta}^{(2)}(\omega,0) \approx \chi_{\gamma\alpha\beta}^{(2)}(\omega,-\omega) \approx \chi_{\gamma\alpha\beta}^{(2)}(\omega,\omega) \approx \chi_{\gamma\alpha\beta}^{(2)}(0,0)$, \ie the off-resonant susceptibility is approximately frequency-independent and universal for all quadratic processes of a given material. Moreover, Kleinman symmetry is seen to be numerically confirmed in the low frequency limit for monolayer Ge$_2$Se$_2$ in Fig.~\ref{fig:spectra}(b), where we find $\chi_{xyy}^{(2)}(0, 0)=\chi_{yxy}^{(2)}(0,0)=\chi_{yyx}^{(2)}(0,0)$ and $\chi_{xzz}^{(2)}(0,0)=\chi_{zxz}^{(2)}(0,0)=\chi_{zzx}^{(2)}(0,0)$.


As mentioned in the introduction, SHG has been widely utilized to characterize crystal symmetries  \cite{Li2013, Wu2014}. Usually, the types of information obtained from NLO responses are unique and inaccessible via other spectroscopic techniques such as absorption spectra. In rotational SHG spectroscopy, linearly polarized light illuminates the material and the polarization-dependent SHG signal is collected. In particular, by inspecting the angular dependence of the SHG response, one can determine the crystal orientation. In Figs.~\ref{fig:spectra}(c) and \ref{fig:spectra}(d), we have plotted the angular dependence of the SHG intensity at normal incidence and frequency $\hbar\omega=E_g/2$ for parallel (blue) and perpendicular (red) polarizations (relative to the incident electric field) for monolayer WS$_2$ and Ge$_2$Se$_2$, respectively. These polarization maps clearly show the underlying symmetries of both crystals. For monolayer WS$_2$, a six-fold rotational symmetry is observed (a signature of crystals with point group D$_{3h}$), which is due to the three-fold rotational symmetry in combination with the additional mirror symmetry of the structure. Note that the SHG electric field (not the intensity) exhibit only a three-fold rotational symmetry, in accordance with the lattice symmetry. Similarly, the two-fold rotational symmetry of Ge$_2$Se$_2$ reveals its low-symmetry structure, and can be understood by analyzing the crystal symmetries. The crystal orientation can be identified using these polar maps, \eg the zigzag and armchair directions in monolayer WS$_2$ are simply aligned along the lobes of parallel and perpendicular responses, respectively. We have provided similar polar graphs for all 375 materials in Supplementary Figs.~2-376 as well as in the C2DB website \cite{C2DB}. These plots should be useful for determination of 2D crystal orientations.


In practice, accurate measurement of NLO properties of 2D materials is highly challenging, and often large variations in the experimental values for the nonlinear susceptibilities of 2D systems can be observed \cite{Autere2018}. This is due to a variety of reasons including the quality of the 2D material, \eg presence of defects, strain, substrate interactions, etc., as well as the employed measurement techniques (\eg most traditional methods for characterizing bulk crystals are not suitable for 2D systems) \cite{Autere2018}. A database of reliable computational reference data could help clarify this situation. For example, in Ref.~\cite{Malard2013} the quadratic susceptibility for monolayer MoS$_2$ was reported to be in the range of $\approx 0.1$ to $0.8$ nm$^2$/V for pump energies $\hbar\omega=1.2$ to 1.6 eV, which compares reasonably with our calculated values of 0.15 to 0.7 nm$^2$/V for the same energy range. On the other hand, the value of 65 nm$^2$/V (assuming 0.65 nm thickness for monolayer) at pump photon energy of $\hbar\omega=1.53$ eV reported in Ref.~\cite{Kumar2013}, deviates by more than two orders of magnitude from our result strongly questioning the validity of the experiment. In Ref.~\cite{Janisch2014}, the quadratic susceptibility of monolayer WS$_2$ was reported to be 5.8 nm$^2$/V at $\hbar\omega=1.2$, which should be compared to our computed value of 0.53 nm$^2$/V. A similar order of magnitude disagreement between theory and experiment was found in Ref.~\cite{Janisch2014}, and may be attributed to a combination of several effects including neglect of excitons and spin-orbit coupling in the calculations, material imperfections, and other experimental uncertainties. Note that excitonic modifications of the response may be greater near resonances than at the DC limit \cite{Trolle2015}. A final example is monolayer Ga$_2$Se$_2$ (with space group 187), for which the quadratic susceptibility was reported to be 0.58, 1.41, and 1.99 nm$^2$/V (assuming a thickness of 0.83 nm) at photon energies of $\hbar\omega=\{0.77,0.92,1.02\}$ eV, respectively \cite{Zhou2015}. Our calculated values for the same frequency range are 0.15-0.25 nm$^2$/V and thus within a factor of five of the experiments. As these examples illustrate, the data provided by our computations should be useful for validation and interpretation of NLO experiments on 2D materials -- at least at a qualitative (\eg polarization dependence) and semi-quantitative (order of magnitude) level.

\begin{figure}[t]
	\centering
	\includegraphics[width=0.9\textwidth]{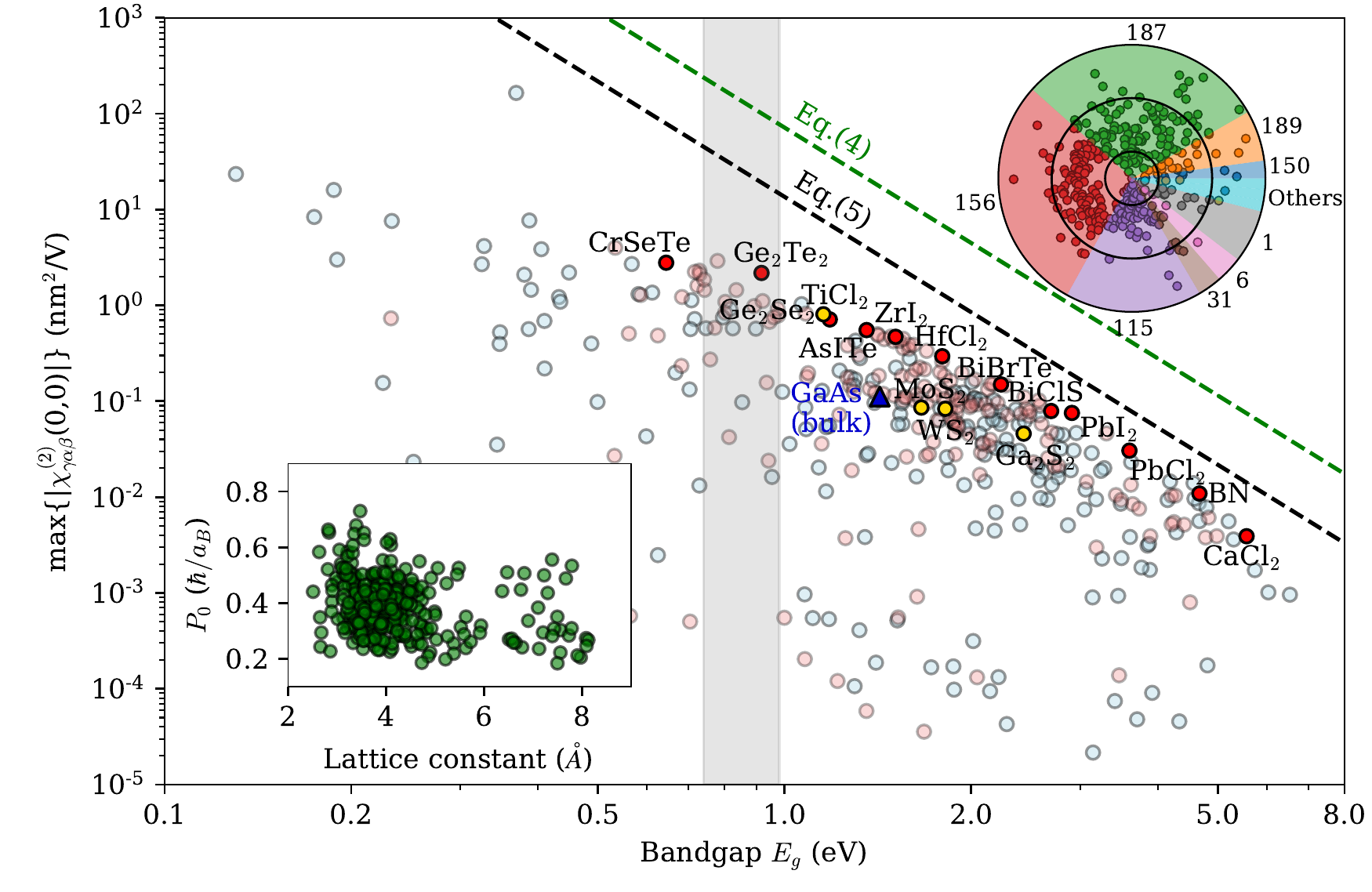}
	\caption[Global]{\textbf{Quadratic response and bandgaps of 2D materials.} Scatter plot (double log scale) of the non-resonant quadratic susceptibility $|\chi_{\gamma\alpha\beta}^{(2)}(0,0)|$ versus bandgap for 375 monolayers. For crystals with several independent tensor elements, the largest element is shown. The light red circles indicate materials, which are predicted to be dynamically and thermodynamically stable (see text), while light blue materials are only marginally stable. Previously studied monolayers with high nonlinearity are shown in yellow, whereas novel promising materials for nonlinear applications are indicated by dark red. For comparison, we also show the experimental non-resonant quadratic susceptibility and bandgap of bulk GaAs as a blue triangle (transformed to sheet susceptibility by multiplying with a thickness of 6.5 {\AA}). The dashed green line indicates the upper limit calculated from Eq.~(\ref{eq:upperlimit0}) by setting $P_0=\hbar/a_B$ and $A_\mathrm{uc}=6.25$ {\AA$^2$} ($a_B=0.529$ {\AA} denotes the Bohr radius). The dashed black line marks the upper limit obtained from Eq.~(\ref{eq:upperlimit}) by setting $E_0=0.2$ eV and $\Xi=1$ (an effective thickness of 6.5 {\AA} is assumed to transform to sheet susceptibility). The gray region is the telecommunication wavelength range from 1260 to 1625 nm. The bottom inset displays the maximum momentum matrix element between the lowest and all higher conduction bands versus the lattice constant for 375 monolayers. The top inset shows the distribution of $\log|\chi_{\gamma\alpha\beta}^{(2)}(0,0)|$ among various space groups, and the three black circles mark $\{0.01, 1, 100\}$ nm$^2$/V from inside to outside.}  
	\label{fig:global}
\end{figure}

%

\subsection*{Theoretical Upper Limit of Quadratic Susceptibility} 
In this section, we investigate the maximum quadratic nonlinearity that is allowed by quantum mechanics, and we clarify the significant role of the bandgap on the theoretical upper bound. As discussed above, for many opto-electronic device applications, the non-resonant nonlinear susceptibility is of great importance. Therefore, we focus on the static quadratic susceptibility as a primary descriptor of optical nonlinearity. Figure~\ref{fig:global} illustrates the non-resonant quadratic susceptibility versus bandgap for all 375 semiconductors studied. The distribution of the data according to crystal space group is shown in the upper right corner. Note that for materials with several independent tensor elements, we only show the largest one.   
Clearly, the quadratic susceptibility tends to be larger for materials with smaller bandgap. This has already been noticed for the shift current generation in crystals \cite{Cook2017, Tan2019} and for the nonlinear response of molecules \cite{Kuzyk2000}. Moreover, for a given bandgap, an upper limit for the non-resonant susceptibility is observed, \ie there is an apparent fundamental limit to the nonlinear susceptibility. In fact, the question of ``what is the largest non-resonant nonlinear susceptibility allowed by quantum mechanics'' has been asked before for molecular systems \cite{Kuzyk2000}. This question can be answered by employing the Thomas-Reiche-Kuhn (TRK) sum rule, which lead to an infinite set of equations relating the transition moments and energies \cite{Mossman2016}. By truncating the Hilbert space to three states, it has been shown that the maximum quadratic susceptibility (hyperpolarizability) of molecules scales as $(E_\mathrm{LUMO}-E_\mathrm{HOMO})^{-3.5}$ \cite{Kuzyk2000}, where $E_\mathrm{HOMO}$ and $E_\mathrm{LUMO}$ are energies of highest occupied and lowest unoccupied molecular orbitals, respectively. Presumably, this model may also be applied to crystals with huge excitonic effects, where the lowest excitons dominate and, hence, constitute an appropriately truncated Hilbert space. However, for crystals treated within the IPA, a continuum of electronic states (bands) exists, which modifies the TRK sum rule and, thereby, the upper bound of quadratic susceptibility.
Below, we derive an upper limit for the non-resonant quadratic response of a semiconductor using two generic models: a three-band model or a two-band model. For simplicity, we only focus on the diagonal element of the susceptibility tensor here, \ie $\gamma=\alpha=\beta$. 



\begin{figure}[p]
	\centering
	\includegraphics[width=0.9\textwidth]{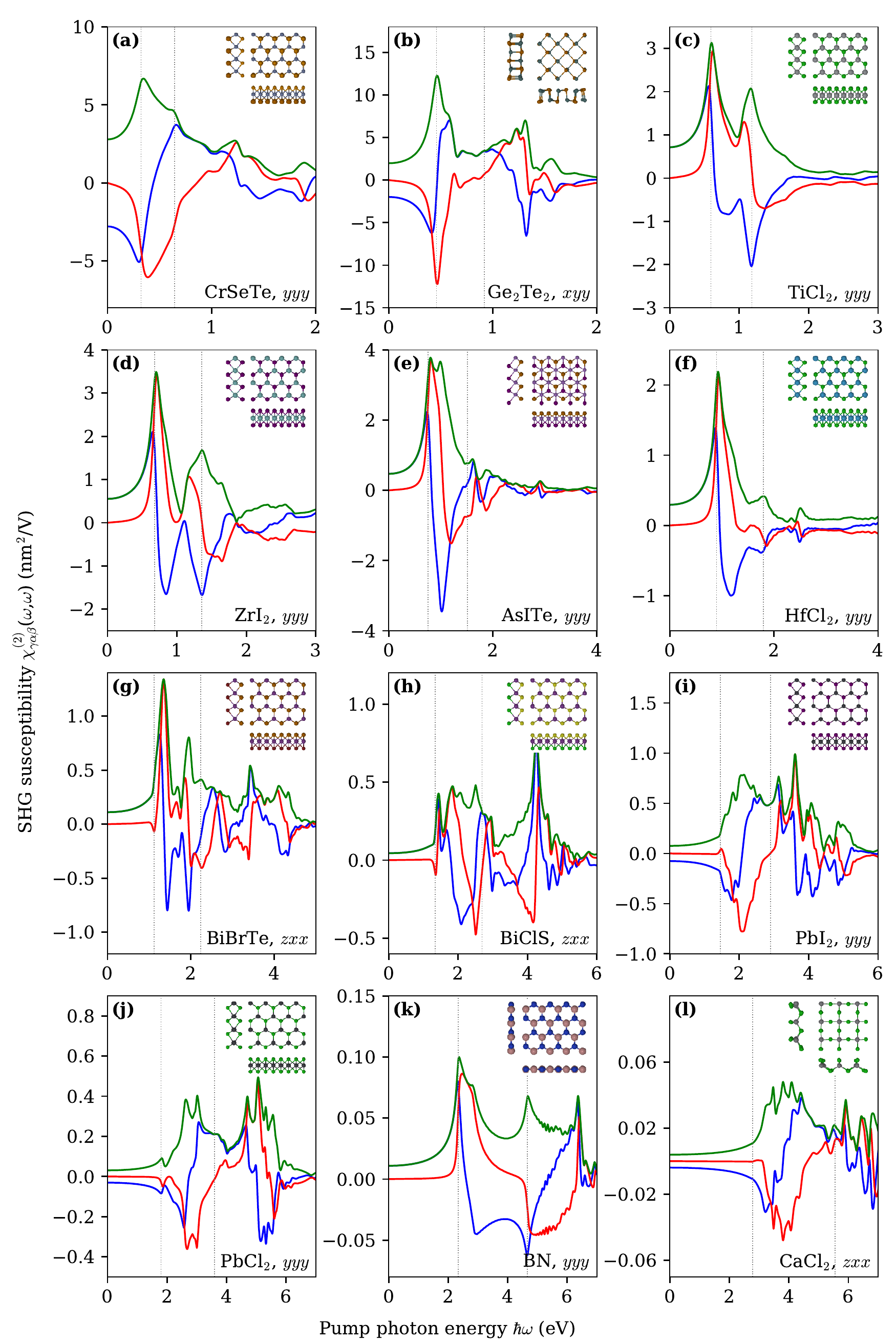}
	\caption[Candidate materials]{\textbf{SHG spectra of 12 monolayers with large NLO response.} Real part (blue), imaginary part (red), and absolute value (green) of $\chi_{\gamma \alpha \beta}^{(2)}(\omega,\omega)$ shown versus pump photon energy for 12 monolayers sorted according to bandgap (low to high) from (a) to (l). For materials with several independent tensor elements, only the component with the largest DC value is shown here (indicated in legends). Vertical dotted lines from left to right mark $\hbar\omega=E_g/2$ and $\hbar\omega=E_g$, respectively. For each material, the crystal structure is shown in the inset including top view and cross sectional views. In all top views, the $x$- and $y$-directions are along the horizontal and vertical directions, respectively.} 
	\label{fig:candidates}
\end{figure}


The upper bounds of both linear and NLO responses originate from the fact that transition dipole matrix elements in a quantum system cannot be arbitrarily large. To understand the implications of such restrictions, we first consider the linear susceptibility of a semiconducting crystal. As demonstrated in the method section, one can derive an upper bound for the DC limit of the diagonal linear susceptibility of a cold semiconductor given by
\begin{align}
\label{eq:linearlimit}
|\chi_{\alpha\alpha}^{(1)}(0)| \leq \frac{2e^2\hbar^2}{\epsilon_0 m V_\mathrm{uc}} \frac{1}{E_g^2} \, ,
\end{align}
where $V_\mathrm{uc}$ is the unit cell volume. We have plotted the non-resonant linear susceptibility versus bandgap for all 375 semiconductors in Supplementary Fig.~1, and demonstrate the validity of Eq.~(\ref{eq:linearlimit}). The bound applies under very general circumstances and highlights the significant role of the bandgap. Moreover, it agrees with previous results for the linear polarizability of molecules, where the upper bound scales as $(E_\mathrm{LUMO}-E_\mathrm{HOMO})^{-2}$ \cite{Mossman2016}. 

To derive the upper bound of the quadratic susceptibility, an additional assumption regarding the number of bands is required. Hence, we first consider a three-band system with one valence and two conduction bands. As shown in the method section, after some algebraic manipulation, the upper limit for the non-resonant quadratic response in the three-band model reads
\begin{align}
\label{eq:upperlimit0}
|\chi_{\alpha\alpha\alpha}^{(2)}(0,0)| \leq \frac{3e^3\hbar^3}{\epsilon_0m^2V_\mathrm{uc}} \frac{P_0}{E_g^4} \, . 
\end{align} 
Here, $P_0$ is the maximum magnitude of the momentum matrix element between the two conduction bands over the Brillouin zone (BZ). An identical upper limit can be obtained in a three-band system with two valence bands, with $P_0$ the maximum momentum matrix element between the two valence bands. In addition, well-defined sheet susceptibilities for 2D materials are obtained by replacing $V_\mathrm{uc}$ with the unit cell area $A_\mathrm{uc}$ in both Eqs.~(\ref{eq:linearlimit}) and (\ref{eq:upperlimit0}). We have collected and plotted the value of $P_0$ for all 375 semiconductors in Fig.~\ref{fig:global}, which clearly exhibits an upper limit on the order of $\hbar/a_B$ with Bohr radius $a_B$. This limit is anticipated since the momentum matrix element for a crystal should be roughly $2\pi\hbar/a$ with the lattice constant $a\approx$2.5-8 {\AA}. By setting the value of $P_0=\hbar/a_B$, the green dashed line in Fig.~\ref{fig:global} is obtained. Note that one can expect that, generally, smaller unit cells are beneficial for realizing larger quadratic response, since $P_0$ tends to increase while $V_\mathrm{uc}$ decreases. 
 
The above-mentioned scaling law for the upper limit of the quadratic response can be derived from a different approach based on a generic two-band Hamiltonian  $\hat{H}(\va{k})=\va{h}(\va{k})\vdot\boldsymbol{\sigma}$, where $\boldsymbol{\sigma}=(\sigma_x,\sigma_y,\sigma_z)$ denotes the vector of Pauli matrices and $\va{h}=(h_x,h_y,h_z)$ are general periodic functions of the 2D wave vector. We assume that the strength of the hopping between atomic sites decreases exponentially with the distance, which is often a reasonable approximation \cite{Tan2019}. With this assumptions and after some simple algebra (see the method section), an upper limit for the DC value of the quadratic susceptibility in two-band model is obtained as
\begin{align}
\label{eq:upperlimit}
|\chi_{\alpha\alpha\alpha}^{(2)}(0,0)| \leq \frac{24e^3}{\epsilon_0} \frac{E_0^2 \Xi}{E_g^4} \, . 
\end{align}  
Here, $E_0$ is an energy that depends only weakly on the bandgap. In addition, $\Xi$ denotes a dimensionless quantity determined by the lattice structure, measurement direction and hopping range (\ie how fast the hopping parameter decreases with distance), see Ref.~\cite{Tan2019} for its general form. $\Xi$ tends to increase for highly asymmetric unit cells (Bravais lattice) and can vary by several orders of magnitude depending on the hopping range. In Ref.~\cite{Tan2019}, the values $\Xi=1$ and $E_0=0.2$ eV were shown to provide an upper bound for the frequency-integrated shift current for a large number of bulk crystals and, therefore, those values are adopted in the present work as well. 
By assuming an effective thickness of 6.5 {\AA} for monolayers, an upper limit shown by the black dashed line in Fig.~\ref{fig:global} is obtained. Note that higher bands are indirectly included in the two band model via the generalized derivative and its sum rule Eq.~(\ref{eq:sumrule}), \cf method section. Therefore, these two seemingly different approaches may lead to similar conclusions. Nevertheless, it is remarkable that both approaches predict an upper limit proportional to $E_g^{-4}$. As noted above, this scaling law is different from the one observed in molecular systems. 


\subsection*{Candidate Materials for 2D Nonlinear Optical Applications} 
In this section, we present the results of our systematic screening of NLO data to identify promising 2D materials with large nonlinear response. The findings are summarized in Fig.~\ref{fig:global}, which provides a handy guide for identifying promising materials with large optical nonlinearities suited for different applications (often dictated by the bandgap). The materials highlighted with a red circle represent monolayers with maximal non-resonant quadratic susceptibility for their bandgap. The highlighted materials are all dynamically stable according to the C2DB, \ie they have no imaginary phonons at the center or corners of the Brillouin zone, and they lie within 0.2 eV/atom of the convex hull \cite{Haastrup2018} meaning that they are among the most thermodynamically stable crystals with the given composition. If a material displays a large DC quadratic susceptibility, all its nonlinearities such as SHG, electro-optic or shift current responses are expected to be large. In Fig.~\ref{fig:candidates}, we plot the SHG spectra for the materials highlighted with red in Fig.~\ref{fig:global}. These materials span several different crystal classes and chemical compositions, and all display stronger nonlinear response than any other 2D material with similar bandgap. For instance, the  quadratic response of monolayer Ge$_2$Te$_2$ is roughly 3 times that of monolayer Ge$_2$Se$_2$ (see Fig.~\ref{fig:spectra}(b)) while their bandgaps differ by only 0.3 eV. Similarly, monolayer HfCl$_2$ possesses a quadratic response that is approximately 10 times larger than that of monolayer WS$_2$ (see Fig.~\ref{fig:spectra}(a)). For comparison purposes, we include the experimental quadratic 
susceptibility of bulk GaAs at a photon energy of $\hbar\omega=0.12$ eV \cite{Eyres2001} in Fig.~\ref{fig:global} as converted into its equivalent sheet susceptibility (assuming 6.5 {\AA} layer thickness similar to MoS$_2$). Bulk GaAs with a zinc blende crystal structure is known to possess an unusually large quadratic response \cite{Boyd2008}. Nonetheless, many of the 2D materials studied here exhibit stronger quadratic response than bulk GaAs. This clearly testifies to the usefulness of 2D materials in nonlinear optics. As a validity check, we have computed the quadratic susceptibility of bulk GaAs with a scissor-corrected bandgap, and find agreement with the experiment within 10 percent.

\begin{figure}[t]
	\centering
	\includegraphics[width=0.97\textwidth]{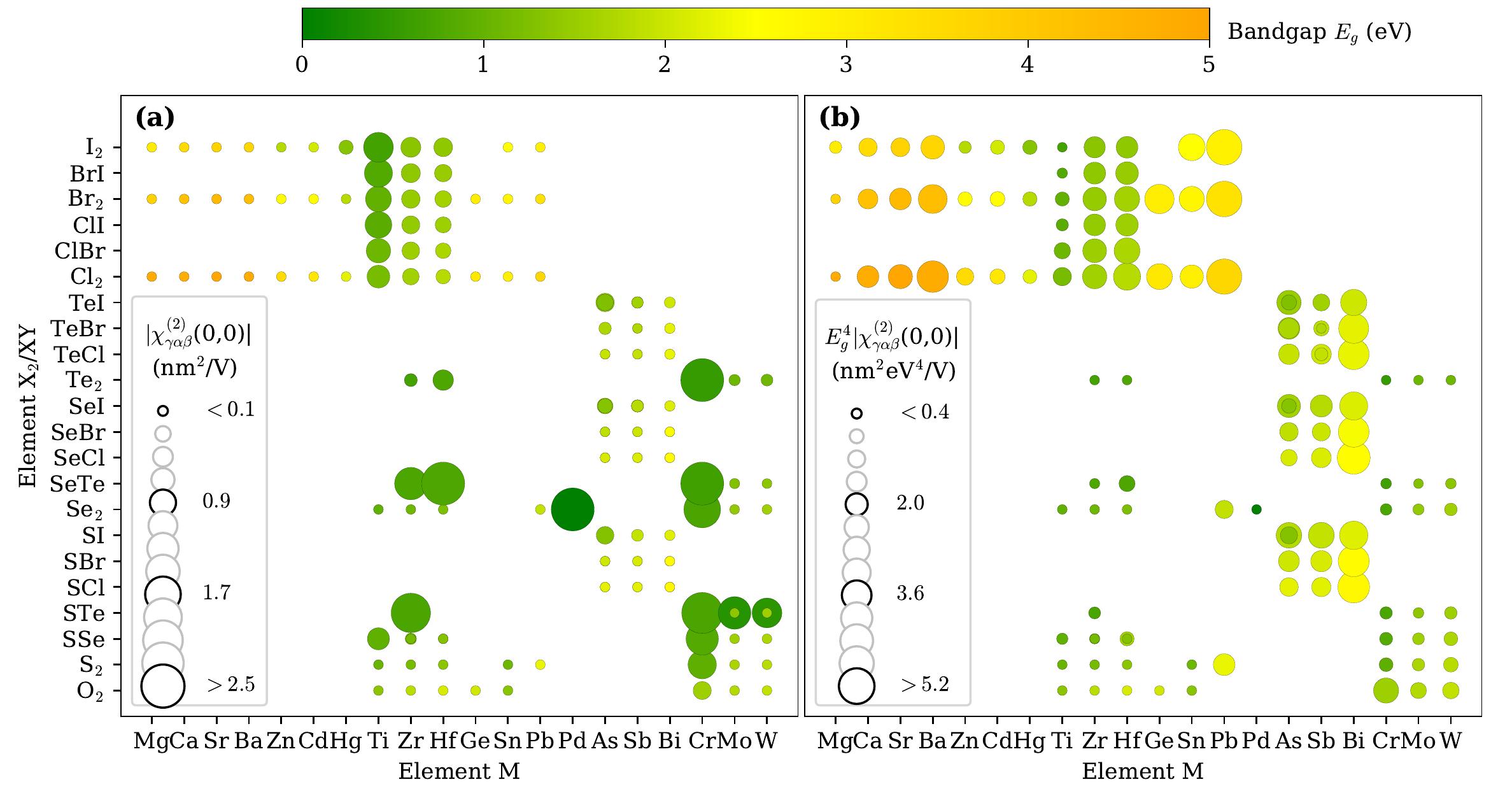}
	\caption[MXY amp]{\textbf{Non-resonant quadratic susceptibility of TMDs and Janus structures.} Plot of (a) $|\chi_{\gamma\alpha\beta}^{(2)}(0,0)|$ and (b) $E_g^4|\chi_{\gamma\alpha\beta}^{(2)}(0,0)|$ for a grid of possible monolayers in the form of MXY (Y can by identical to X) with space group 187 or 156 corresponding to TMDs or their Janus variants. The bandgap is color coded according to the color bar. For Janus structures with several independent tensor elements only the component with the largest DC value is shown.  } 
	\label{fig:mxy}
\end{figure}

Our NLO data set can be used to explore correlations between basic material properties and nonlinear responses. As an example, Fig.~\ref{fig:mxy}(a) illustrates the non-resonant quadratic susceptibility of more than 150 different monolayers with either space group 187 and chemical formula MX$_2$ or space group 156 and chemical formula MXY versus chemical constituents. These two classes of monolayer materials cover the popular TMDs and their recently introduced Janus flavors \cite{RiisJensen2019}, respectively. The empty entries in the matrix correspond to materials that either do not meet our stability criteria or do not exhibit a finite bandgap. Several trends can be extracted from the plot, highligting the trade-off between large nonlinear response and bandgap, \ie transparency range. 
For instance, materials including Cr, Ti or Te consistently exhibit quite strong nonlinear response but typically at smaller bandgaps, \eg CrSe$_2$, CrSeTe or TiCl$_2$. To appropriately consider the effect of bandgap, we have also re-plotted Fig.~\ref{fig:mxy}(a) by normalizing the data with the bandgap ($E_g^4 |\chi_{\gamma \alpha \beta}^{(2)}(0,0)|$) in Fig.~\ref{fig:mxy}(b). It can be observed that monolayers with halogens (Cl, Br and I) or Bi exhibit strong quadratic responses considering the material bandgap. This is also confirmed in Fig.~\ref{fig:candidates}, where several promising materials include halogens and/or Bi. Similar analyses can be performed for other material classes or properties to extract further trends and correlations from our data set.

\section*{Discussion}
Our work has identified several promising 2D materials for a wide range of opto-electronic applications. It is hoped that by using the DC quadratic response as a descriptor, appropriate materials for operation at finite frequencies can be reliably identified. Moreover, the full SHG spectra collected in Supplementary Information provide detailed predictions of dynamical effects. Although we have focused on the quadratic nonlinearity and only reported SHG spectra, the present approach can be readily extended to other quadratic responses such as shift currents or higher-order nonlinearities including third-harmonic generation or optical Kerr effects, albeit with higher computational cost. In addition, the present methodology could potentially target a much broader range of materials beyond the non-magnetic monolayer class. An obvious extension would be multilayers and van der Waals heterostructures that are attractive for realizing exotic NLO angular response patterns \cite{He2019} or tunable nonlinearities \cite{Hsu2014}. Similarly, including magnetic materials, such as the peculiar ferromagnet CrI$_3$, is an intriguing avenue for ab initio NLO response calculation. Regarding the computational methodology, we have neglected excitonic effects and self-energy corrections in the present work due to their computational cost. Many-body interactions unquestionably affect the dynamical NLO response of 2D materials, as evident by strong exciton binding. Hence, an immediate application of our screening study is to point to highly interesting materials, which might then be investigated more accurately within the GW and BSE frameworks. We caution, however, that the computational cost of excitonic NLO response calculations is substantially greater than in the linear case. It is conceivable that additional theoretical developments in this highly active research field could provide much needed computational tools for this purpose. Finally, other less-important effects such as local fields, substrate interactions, defects, and coupling to phonons could be investigated in future studies.

A separate significant result of the present work is the establishment of a quantitative upper theoretical limit for the quadratic optical response. Our high-throughput screening of 375 2D semiconductors supports the limiting scaling $\mathrm{max}(|\chi_{\gamma\alpha\beta}^{(2)}|)\propto E_g^{-4}$ derived for the maximum possible optical nonlinearity of a any material. In this regard, it is intriguing that our best 2D candidates fall short of the upper limit by factors of only 2-3 or an order of magnitude using Eqs.~(\ref{eq:upperlimit}) or (\ref{eq:upperlimit0}), respectively. These particular 2D materials have nonlinearities significantly larger than all previously-reported monolayers with similar bandgaps. For instance, stable monolayer HfCl$_2$ (lying only 0.01 eV/atom above the convex hull) with a bandgap in the near-infrared exhibits 10 times stronger nonlinear response than that of monolayer WS$_2$. Similarly, monolayer Ge$_2$Te$_2$ with a bandgap in the telecommunication range possesses a huge quadratic response. Combined with its ferroelasticity and ferroelectricity properties, this makes it highly attractive for various opto-electronic applications. Importantly, these 2D materials exhibit stronger nonlinear responses than known bulk systems. One may also anticipate that the inclusion of excitonic effects may further boost the predicted NLO response of 2D materials compared to bulk systems as a consequence of the pronounced many-body effects in the former group \cite{Taghizadeh2018}.

As an important representative of our results, we have provided a comprehensive library of SHG spectra and their polarization dependence, freely available for browsing or download from the C2DB website \cite{C2DB}. The C2DB by now is a unique and powerful resource for optical characterization of 2D materials, since it provides access to linear and NLO responses as well as infrared and Raman spectra. Our library can be used as a valuable reference for NLO experiments on 2D materials, particularly for validation and interpretation of experimental results \eg the polarization dependence. The database also facilitates design of NLO devices such as frequency converters or electro-optic modulators. Here, our correlation of nonlinearity with bandgap should be useful for ensuring that undesirable levels of absorption can be avoided at the operation wavelength. Apart from being a useful reference for 2D materials research, our library can be used to train machine learning algorithms for materials science problems \cite{Schmidt2019}. Similarly to recent work on prediction of linear optical spectra for molecules \cite{Ghosh2019}, our database may enable prediction of NLO spectra directly from the atomic structure and, in turn, autonomous (in situ) characterization of materials \cite{Taghizadeh2020}.



\section*{Methods \label{sec:Methods}}


\subsection*{Theory}
In the IPA, the Hamiltonian of an electron interacting with an external electromagnetic field takes the form $\hat{H}=\hat{H}_0+\hat{H}_\mathrm{int}$, where $\hat{H}_0$ is the unperturbed Hamiltonian and $\hat{H}_\mathrm{int}$ describes the electron-light interaction. 
Within DFT, $\hat{H}_0$ is the Kohn-Sham Hamiltonian with the Bloch eigenstates, \ie $\hat{H}_0\ket{n\va{k}}=\varepsilon_{n\va{k}}\ket{n\va{k}}$, labeled by a band index $n$ and wavevector $\va{k}$, while $\varepsilon_{n\va{k}}\equiv\hbar\omega_{n\va{k}}$ denotes the single-particle energies. Note that, to simplify the notation, we frequently suppress the explicit wavevector dependence of various quantities. We restrict ourselves here to non-magnetic systems and neglect spin-orbit coupling, which leads to spin-degenerate states. 
In the long-wavelength limit (\ie neglecting the spatial variation of the external field), the interaction Hamiltonian can be written in either velocity or length gauge. 
These two approaches are formally equivalent (due to gauge freedom) but may lead to different results in practical calculations because of various approximations \cite{Taghizadeh2017, Taghizadeh2018}. In this work, we choose to work in the length gauge, since, for a finite basis set, it produces more accurate results compared to the velocity gauge \cite{Taghizadeh2017}. Nonetheless, we have verified for all materials that the spectra calculated using the velocity gauge agree with the length gauge to an excellent degree and, hence, that gauge invariance is satisfied numerically. The interaction Hamiltonian in the length gauge is given by $\hat{H}_\mathrm{int}(t) \equiv e\field{F}(t)\vdot\rop$, where $\rop$ denotes the position operator.

In the Heisenberg picture, the above-mentioned Hamiltonians lead to the usual Liouville equation of motion for the reduced density matrix, which we solve perturbatively to determine the $N$th order density matrix \cite{Taghizadeh2017, Ventura2017, Hipolito2018}. Since the position operator is ill-defined in periodic systems it requires special attention. A rigorous treatment of the length gauge optical response proceeds by formally separating the interband and intraband parts of the position operator according to $\rop=\rop^{(e)}+\rop^{(i)}$ \cite{Sipe2000, Taghizadeh2017}, where $\mel{n\va{k}}{\rop^{(e)}}{m\va{k}'} \equiv (1-\delta_{nm}) \delta_{\va{k}\va{k}'} \boldsymbol{\Omega}_{nm}$ and $\mel{n\va{k}}{\rop^{(i)}}{m\va{k}'}\equiv\delta_{nm} (\boldsymbol{\Omega}_{nn}+i\nabla_\va{k}) \delta_{\va{k}\va{k}'}$ \cite{Sipe2000}. 
Here, $\delta$ denotes the Kronecker delta and the generalized Berry connection $\boldsymbol{\Omega}_{nm}$ is defined as
\begin{align}
\boldsymbol{\Omega}_{nm} \equiv \frac{i}{V_\mathrm{uc}} \int_\mathrm{uc} u_{n\va{k}}^*(\va{r}) \nabla_{\va{k}} u_{m\va{k}}(\va{r}) \mathrm{d}^3{\va{r}} \, , \
\end{align}
where $u_{n\va{k}}(\va{r}) \equiv \sqrt{V}\exp(-i\va{k}\vdot\va{r}) \braket{\va{r}}{n\va{k}}$ denotes the periodic part of the Bloch state, and $V$ is the crystal volume. Note that the interband Berry connections are simply related to the momentum matrix elements via the commutator relation $[\hat{H}_0,\rop^{(e)}]=\hbar\pop/(im)$. This leads to $\boldsymbol{\Omega}_{nm}= \va{p}_{nm}/(im\omega_{nm})$ for $n\neq m$, with $\omega_{nm}\equiv\omega_{n\va{k}}-\omega_{m\va{k}}$ and momentum matrix elements $\va{p}_{nm}\equiv\mel{n\va{k}}{\pop}{m\va{k}}$. The problematic intraband part of the position operator leads to the appearance of the so-called generalized derivative in response functions \cite{Aversa1995}. For any simple operator $\hat{O}$ (diagonal in $\va{k}$), the commutator with the intraband position operator is, again, a simple operator and satisfies the relation $\mel{n\va{k}}{[r_\alpha^{(i)},\hat{O}]}{m\va{k}'}\equiv i\delta_{\va{k}\va{k}'} (O_{nm})_{;\alpha}$. Here, the generalized derivative is defined as $(O_{nm})_{;\alpha} \equiv \partial_{\alpha} O_{nm} - i(\Omega_{nn}^\alpha-\Omega_{mm}^\alpha) O_{nm}$ with $\partial_{\alpha}\equiv\partial/\partial k_\alpha$.

Upon obtaining the density matrix $\rho(t)$, the expectation value of the current density operator $\jop=-eg\pop/(mV)$ is determined straightforwardly as $\field{J}(t)\equiv\mathrm{Tr}[\jop \rho(t)]$, where $g = 2$ accounts for the spin-degeneracy. The optical polarization is then related to the current density via $\field{J}(t)=\partial \field{P}(t)/\partial t$, which leads to an expression for the susceptibility tensor. In the original form, the derived expression is not suitable for numerical implementation due to appearance of double poles and, hence, should be modified by partial-fraction decomposition and imposing time-reversal symmetry \cite{Sipe2000}. In general, the quadratic susceptibility in the length gauge consists of four distinct terms: one purely interband, two mixed inter-intraband, and one purely intraband contributions \cite{Aversa1995}. For a cold, intrinsic semiconductor, the purely intraband and one of the mixed inter-intraband parts vanish identically.
In turn, the total quadratic response is reduced to a purely interband term $\chi_{\gamma\alpha\beta}^{(2e)}(\omega_1,\omega_2)$ and a mixed term  $\chi_{\gamma\alpha\beta}^{(2i)}(\omega_1,\omega_2)$ that read \cite{Sipe2000}
\begin{align}
	%
	\label{eq:chiinter}
	\chi_{\gamma\alpha\beta}^{(2e)}(\omega_1,\omega_2) &\equiv C_0 \sideset{}{'}\sum_{\va{k},nml} \dfrac{r_{nm}^\gamma r_{ml}^\alpha r_{ln}^\beta }{s_1\omega_{ln}-s_2\omega_{ml}} \bigg( \frac{f_{nm}}{\omega_s-\omega_{mn}} - \frac{s_2f_{nl}}{\omega_2-\omega_{ln}} + \frac{s_1f_{ml}}{\omega_1-\omega_{ml}} \bigg) + (\alpha,\omega_1\rightleftarrows\beta,\omega_2) \, , \\
	\label{eq:chiintra}
	\chi_{\gamma\alpha\beta}^{(2i)}(\omega_1,\omega_2) &\equiv iC_0 \sideset{}{'}\sum_{\va{k},nm} \frac{f_{nm} r_{nm}^\gamma}{\omega_s-\omega_{mn}} \bigg[ \frac{(r_{mn}^\alpha)_{;\beta}}{s_2\omega_{mn}} - \frac{r_{mn}^\alpha \Delta_{mn}^\beta}{s_2^2\omega_{mn}^2} \bigg] 
	+\frac{iC_0}{2} \sideset{}{'}\sum_{\va{k},nm} \frac{f_{nm} r_{mn}^\alpha}{\omega_1-\omega_{mn}}  \bigg[ \frac{s_1(r_{nm}^\gamma)_{;\beta}}{s_2\omega_{mn}} + \frac{s_1^2 r_{nm}^\gamma \Delta_{mn}^\beta}{s_2^2\omega_{mn}^2} \bigg] \nonumber \\
	& -\frac{iC_0}{\omega_s} \sideset{}{'}\sum_{\va{k},nm} \frac{f_{nm} r_{mn}^\beta (r_{nm}^\alpha)_{;\gamma}}{\omega_2-\omega_{mn}}
	-\frac{iC_0}{\omega_s^2} \sideset{}{'}\sum_{\va{k},nm} \frac{f_{nm} r_{nm}^\alpha r_{mn}^\beta \Delta_{mn}^\gamma}{\omega_2-\omega_{mn}} + (\alpha,\omega_1\rightleftarrows\beta,\omega_2) \, .
\end{align}
Here, $C_0\equiv ge^3/(2\hbar^2\epsilon_0V)$, $\omega_s\equiv\omega_1+\omega_2$, $s_i\equiv\omega_i/\omega_s$ for $i=1,2$, $\Delta_{mn}^\alpha\equiv (p_{mm}^\alpha-p_{nn}^\alpha)/m$ denotes the velocity difference between bands $n$ and $m$, and $r_{nm}^\alpha=\Omega_{nm}^\alpha$ is the interband ($n \neq m$) position matrix element. In addition, $f_{nm}\equiv f_n-f_m$, where $f_n\equiv(1+\exp[(\varepsilon_n-\mu)/k_BT])^{-1}$ is the Fermi–Dirac distribution with chemical potential $\mu$ at temperature $T$. The primed summation signs indicate omission of terms with two or more identical indices. Also, the summation over $\va{k}$ implies an integral over the first BZ, \ie $(2\pi)^3 \sum_\va{k} \rightarrow V \int_\mathrm{BZ} \mathrm{d}^3{\va{k}}$. Finally, $(\alpha,\omega_1\rightleftarrows\beta,\omega_2)$ indicates that the preceding terms should be added with  $\alpha$ and $\beta$ as well as $\omega_1$ and $\omega_2$ interchanged. We note that the quadratic optical conductivity can readily be derived from the corresponding susceptibility
via $\sigma_{\gamma\alpha\beta}^{(2)}(\omega_1,\omega_2)=-i\epsilon_0(\omega_1+\omega_2)\chi_{\gamma\alpha\beta}^{(2)}(\omega_1,\omega_2)$. In the specific case $\omega_1=-\omega_2=\omega$, the first and second terms in the last line of Eq.~(\ref{eq:chiintra}) are identified with shift and injection currents, respectively \cite{Wang2019}. In the above summations, special care is required at band crossings as discussed in Ref.~\cite{Sipe2000}. However, their contribution is negligible for any sufficiently fine $\va{k}$-mesh, since the volume represented by degenerate points in the BZ approaches zero \cite{Wang2017c}. In our calculations, we discard near-degenerate transitions with energy differences smaller than a chosen threshold to avoid numerical divergences. We have confirmed that varying the energy threshold from 10$^{-5}$ to 10$^{-8}$ eV does not affect the calculated spectra. 

All the ingredients in Eqs.~(\ref{eq:chiinter}) and (\ref{eq:chiintra}) are either available or readily obtainable from a generic DFT calculation, except the generalized derivative. Although the required matrix elements may be computed directly from a finite-difference scheme \cite{Taghizadeh2017, Taghizadeh2018}, care should be taken of the random phases of Bloch states \cite{Sipe2000}. Alternatively, by virtue of the canonical commutation relation $[\hat{r}_\alpha, \hat{p}_\beta]=i\hbar \delta_{\alpha\beta}$ and by separating inter- and intraband parts of the position operator, the generalized derivative (for $n \neq m$) may be evaluated from the sum-rule \cite{Aversa1995},
\begin{align}
\label{eq:sumrule}
(r_{nm}^\beta)_{;\alpha} &= \frac{r_{nm}^\alpha \Delta_{mn}^\beta + r_{nm}^\beta \Delta_{mn}^\alpha }{\omega_{nm}} + \frac{i}{\omega_{nm}} \sum_{l\neq n,m} (\omega_{lm} r_{nl}^\alpha r_{lm}^\beta - \omega_{nl} r_{nl}^\beta r_{lm}^\alpha) \, .
\end{align}
Equation~(\ref{eq:sumrule}) is used in this work to evaluate the required generalized derivatives, with the only limitation that infinite sums are replaced by finite sums over a restricted, but large, number of bands. 
 


\subsection*{Upper Limit of Static Linear Susceptibility}
The general expression for the linear susceptibility of a cold semiconductor in the dipole approximation is given by \cite{Sipe2000}
\begin{align}
\chi_{\gamma\alpha}^{(1)}(\omega) \equiv \frac{ge^2}{\epsilon_0\hbar V} \sideset{}{'}\sum_{\va{k},nm} \frac{f_{nm}r_{nm}^\gamma r_{mn}^\alpha}{\omega_{mn}-\omega} = \frac{2ge^2}{\epsilon_0m^2\hbar V} \sum_{\va{k},cv} \frac{\Re[p_{cv}^\gamma p_{vc}^\alpha]}{\omega_{cv}(\omega_{cv}^2-\omega^2)} \, ,
\end{align}
where, in the second equality, time reversal symmetry has been employed. In addition, for a cold semiconductor, we have $f_{c/v}=\{0,1\}$, where the indices $c$ and $v$ imply conduction and valence bands, respectively. Then, the upper bound for the static limit of the diagonal linear susceptibility is given by 
\begin{align}
\chi_{\alpha\alpha}^{(1)}(0) = \frac{2ge^2}{\epsilon_0m^2\hbar V} \sum_{\va{k},cv} \frac{|p_{cv}^\alpha|^2}{\omega_{cv}^3} \leq \frac{2ge^2\hbar}{\epsilon_0m^2 V E_g^2} \sum_{\va{k},cv} \frac{|p_{cv}^\alpha|^2}{\omega_{cv}} = \frac{2e^2\hbar^2}{\epsilon_0 m V_\mathrm{uc}} \frac{1}{E_g^2} \, .
\end{align}
In the last equality here, we have employed a TRK sum rule for crystals, given by $2\sum_{\va{k},cv} |p_{cv}^\alpha|^2/\omega_{cv} = m\hbar N_e$ with $N_e=V/V_\mathrm{uc}$ as the total number of electrons, see Supplementary Information for the derivation. From this expression, one can confirm that to achieve the maximum static linear susceptibility, it is advantageous to have higher conduction (lower valence) bands as far as possible from the lowest conduction (highest valence) band. Hence, only a two-band system with negligible dispersion for $\omega_{cv}$, \ie $\hbar\omega_{cv}\approx E_g$ over the BZ, may realize the maximum of the linear susceptibility.  

\subsection*{Upper Limit of Static Quadratic Susceptibility}
The general expression for the static quadratic susceptibility can simply be derived by letting both $\omega_1$ and $\omega_2$ approach zero, while keeping $s_1+s_2=1$.  We focus on the diagonal tensor element of the static quadratic susceptibility, for which the expression is given in Supplementary Information section 1.

\textbf{Three-band model:} We consider a three-band system with two conduction ($c,c'$) and one valence ($v$) bands. Without loss of generality, we assume that $c$ is the lowest conduction band, \ie $\hbar\omega_{c'v} \geq \hbar\omega_{cv} \geq E_g$. Hence, the expression for the static quadratic susceptibility is simplified to
\begin{align}
\label{eq:threeband}
	\chi_{\alpha\alpha\alpha}^{(2)}(0,0) = - \frac{6C_0}{m^3} \sum_{\va{k}} \frac{\Im[p_{vc}^\alpha p_{cc'}^\alpha p_{c'v}^\alpha] \omega_{c'c}}{\omega_{cv}^2\omega_{c'v}^2} \Big(\frac{2}{\omega_{cv}^2}+\frac{2}{\omega_{c'v}^2}+ \frac{3}{\omega_{cv}\omega_{c'v}}\Big) = \frac{6C_0}{m^2} \sum_{\va{k}} \frac{\Im[p_{vc}^\alpha (p_{cv}^\alpha)_{;\alpha}]}{\omega_{cv}^4}  f(\delta)  \, ,
\end{align}
where the dimensionless function $f(\delta)$ is defined as $f(\delta) \equiv (1-\delta)^2(2\delta^2+3\delta+2)/(2-\delta)$ with $0 \leq \delta \equiv \omega_{cv}/\omega_{c'v} \leq 1$. In the second equality, the sum rule for the generalized derivative of the momentum matrix element is used to replace $p_{cc'}^\alpha p_{c'v}^\alpha$, \ie $m\omega_{c'v}\omega_{cc'}(p_{cv}^\alpha)_{;\alpha} \approx p_{cc'}^\alpha p_{c'v}^\alpha(\omega_{c'v}-\omega_{cc'})$ \cite{Taghizadeh2017}. Next, we employ the Cauchy–Schwarz inequality to obtain the upper bound of Eq.~(\ref{eq:threeband}) as $|\chi_{\alpha\alpha\alpha}^{(2)}(0,0)|= 6C_0|\Im[A]|/m^2 \leq 6C_0|A|/m^2$, with $A$ defined from
\begin{align}
|A|^2\equiv \Big|\sum_{\va{k}} \frac{p_{vc}^\alpha (p_{cv}^\alpha)_{;\alpha}}{\omega_{cv}^4} f(\delta) \Big|^2 \leq \Bigg( \sum_{\va{k}} \frac{|p_{vc}^\alpha|^2 |f(\delta)|^2}{\omega_{cv}^8} \Bigg) \Bigg( \sum_{\va{k}} |(p_{cv}^\alpha)_{;\alpha}|^2 \Bigg) \, .
\end{align}
Since $0 \leq |f(\delta)| \leq 1$ for any $0\leq \delta \leq 1$ and $|f(\delta)|$ is decreasing monotonically with $\delta$, we replace it by $f(0)=1$ to maximize the first term. Note that the case $\delta\rightarrow0$ corresponds to a situation, where the second conduction band ($c'$) is far from the lowest conduction band ($c$), which has also been shown to maximize the quadratic nonlinearity in molecular systems \cite{Kuzyk2000}. In addition, this is in agreement with the required condition for maximizing the linear susceptibility (see above). Next, one can derive a useful upper bound for momentum matrix elements between the conduction and valence bands, which reads $2\sum_\va{k} |p_{vc}^\alpha|^2/\omega_{cv}^N \leq \hbar^N m N_e/E_g^{N-1}$ for any integer $N\geq 1$. This upper bound is derived using the identity $2\sum_{\va{k},cv} |p_{cv}^\alpha|^2/\omega_{cv} = m\hbar N_e$, as shown in Supplementary section~1. Hence, the upper bound for the first term is $\hbar^8mN_e/2E_g^7$.
Regarding the second term, the generalized derivative is replaced by $m|(p_{cv}^\alpha)_{;\alpha}| \approx |p_{cc'}^\alpha| |p_{c'v}^\alpha| |\omega_{cc'}^{-1}-\omega_{c'v}^{-1}| \leq P_0 |p_{c'v}^\alpha|/\omega_{c'v}$ (see above) with the maximum value of $|p_{cc'}^\alpha|$ as $P_0$, \ie $|p_{cc'}^\alpha| \leq P_0$. Hence, the upper bound of the second term is $\hbar^2P_0^2N_e/(2mE_g)$. Combining the bounds of the first and second terms leads to Eq.~(\ref{eq:upperlimit0}) for the upper limit of the static quadratic susceptibility. 


\textbf{Two-band model:} In a two-band system, the purely interband term vanishes, and only the mixed inter-intraband term survives. Focusing on the diagonal tensor element of this contribution, the static quadratic susceptibility reads
\begin{align}
\label{eq:staticsusceptibility}
	\chi_{\alpha\alpha\alpha}^{(2)}(0,0) = \frac{6e^3}{\epsilon_0 V} \sum_{\va{k}} \frac{\Im[r_{vc}^\alpha (r_{cv}^\alpha)_{;\alpha}]}{\varepsilon_{cv}^2} \, .
\end{align}  
For the generic two-band Hamiltonian, the eigenenergies are given by $\pm|\va{h}|$, \ie $\varepsilon_{cv}=2|\va{h}|$. Next, we take the derivative of $\mel{n\va{k}}{H(\va{k})}{m\va{k}}=\varepsilon_{n\va{k}} \delta_{nm}$ with respect to $k_\alpha$ twice.
After some algebraic manipulations, $\Im[r_{vc}^\alpha (r_{cv}^\alpha)_{;\alpha}]$ reads $4\Im[r_{vc}^\alpha (r_{cv}^\alpha)_{;\alpha}]=-\va{h}\vdot(\va{h}'\times\va{h}'')/|\va{h}|^3$ \cite{Cook2017},
where $\va{h}'$ and $\va{h}''$ denote the first and second derivatives of $\va{h}(\va{k})$ with respect to $k_\alpha$. Now, since $\va{h}(\va{k})$ is periodic in reciprocal space, it can be Fourier transformed. Without loss of generality, $\va{h}(\va{k})$ can be written as  $\va{h}(\va{k})=\sum_{\va{R}} \va{h}_{\va{R}} \exp(i\va{k}\vdot\va{R})$, where $\va{R}$ denotes lattice vectors. The Fourier coefficients, $\va{h}_{\va{R}}$, are basically the hopping parameters in a tight-binding picture, or the hopping amplitudes between Wannier functions. Finally, we use the property $\va{h}_{\va{R}_1}\vdot(\va{h}_{\va{R}_2}\times\va{h}_{\va{R}_2}) \leq |\va{h}_{\va{R}_1}| |\va{h}_{\va{R}_2}| |\va{h}_{\va{R}_2}|$, and assume an exponential decay of $\va{h}_{\va{R}}$ with the magnitude of the lattice vector $|\va{R}|$, \ie $|\va{h}_{\va{R}}|\leq E_0 \exp(-|\boldsymbol{\zeta}\vdot\va{R}|)$ with $\boldsymbol{\zeta}$ a hopping range parameter. Following the approach presented in Ref.~\cite{Tan2019}, the $\va{k}$-integration leads to the $\Xi$ function, and the upper limit of Eq.~(\ref{eq:upperlimit}) is derived in an straightforward manner.


\subsection*{First-principles Calculations}
All DFT calculations were performed with the projector augmented wave code, GPAW \cite{Enkovaara2010}, in combination with the atomic simulation environment (ASE) \cite{HjorthLarsen2017}. The Perdew-Burke-Ernzerhof (PBE) exchange-correlation functional was used \cite{Perdew1996} and the Kohn-Sham orbitals were expanded using a plane wave basis with cut-off energy of 600 eV. The monolayers were repeated periodically in supercells including 15 {\AA} vacuum between layers. Based on convergence tests for the SHG spectrum performed for several materials, a $\va{k}$-mesh of density 30 {\AA} was judged to be sufficient. The number of empty bands included in the sum over bands was set to three times the number of occupied bands. The width of the Fermi-Dirac occupation factor was set to $k_BT=50$ meV in the DFT calculations. Furthermore, a line-shape broadening of $\eta = 50$ meV was used in all spectra. Time-reversal symmetry was imposed in order to reduce the $\va{k}$-integrals to half the BZ. For several 2D crystal classes, we have verified by explicit calculation that the quadratic tensor elements fulfill the expected symmetries, \eg that they all vanish identically for crystals with an inversion center. The calculations were submitted, managed, and received using the MyQueue workflow tool \cite{Mortensen2020}, which is a Python front-end to the batch job scheduler.


\section*{Data Availability}
All calculated second-harmonic spectra are freely available online through the \href{https://cmrdb.fysik.dtu.dk/c2db/}{C2DB website}. Other data is available from the corresponding author upon reasonable request. 


\section*{Code Availability}
GPAW is an open-source DFT Python code based on the projector-augmented wave method and the ASE, which is available at the \href{https://wiki.fysik.dtu.dk/gpaw/}{GPAW website}. The nonlinear code used for generating second-harmonic generation spectra in this work will be available in future releases of GPAW code.




\section*{Acknowledgments}
This work was supported by the Center for Nanostructured Graphene (CNG) under the Danish National Research Foundation (project DNRF103). K. S. T. acknowledges support from the European Research Council (ERC) under the European Union’s Horizon 2020 research and innovation program (Grant No. 773122, LIMA).


\section*{Author Contributions}
A.T. wrote the computer code, carried out the calculations, and validated the data. All authors discussed the results and contributed to writing the manuscript. T.G.P. and K.S.T. supervised the project. 


\section*{Additional information}
\noindent\textbf{Supplementary Information:} accompanies this paper.

\noindent\textbf{Competing interests:} The authors declare no competing interests.


\printbibliography

\end{document}